\documentclass[12pt]{iopart}

\usepackage{iopams}  
\usepackage{graphicx}
\usepackage{dcolumn}% Align table columns on decimal point
\usepackage{bm}% bold math
\usepackage{color}
\definecolor{DarkGreen}{rgb}{0,0.7,0.08} 
\definecolor{Grey}{rgb}{0.5,0.5,0.5}
\definecolor{Red}{rgb}{0.8,0.3,0.3}
\newcommand{\commentpj}[1]{#1}

\newcommand{\changed}[1]{#1}

\begin{document}

\title[]{Quantum de-mixing in binary mixtures of dipolar bosons}

\author{Piyush Jain}
\address{Department of Physics, University of Alberta, Edmonton, Alberta, Canada}
\ead{jain@ualberta.ca}
\author{Massimo Boninsegni}
\address{Department of Physics, University of Alberta, Edmonton, Alberta, Canada}
\ead{m.boninsegni@ualberta.ca}

\begin{abstract}
Quantum Monte Carlo simulations of a two-component Bose mixture of \commentpj{trapped} dipolar atoms of identical masses and dipole moments, provide numerical evidence of de-mixing at low \commentpj{finite temperatures}. De-mixing occurs as a consequence of  quantum statistics, which results in an effective attraction between like bosons. Spatial separation of two components takes place at low temperature with the onset of long exchanges of identical particles, underlying Bose-Einstein condensation of both components. Conversely, at higher temperature the system is miscible due to the entropy of mixing. \commentpj{Exchanges are also found to enhance demixing in the case of mixtures of non-identical and distinguishable species.}
\end{abstract}

%\maketitle

\section{Introduction}

Phase separation or \emph{de-mixing} in multi-component mixtures has been a long standing topic of interest in chemistry and physics. Closely following the achievement of Bose-Einstein condensation in dilute gases \cite{Anderson1995,Davis1995}, there has been considerable interest in the study of binary mixtures of Bose-Einstein condensates \changed{(BECs)}. Of particular note, a two-component \changed{BEC} was first reported in 1997 comprised of two hyperfine states of Rb \cite{Myatt1997}, and then in 2001 using different atomic species (K and Rb) \cite{Modugno2001}. The advantage of these ultracold systems is that the entropy of mixing is small and de-mixing may be easily observed. 

The conditions under which de-mixing occurs in binary BEC mixtures with hard-core repulsion, have been the focus of a number of theoretical works, including mean-field treatments at zero \cite{Ho1996, Ao1998, Timmermans1998, Pu1998, Esry1999, Trippenbach2000} and finite temperature \cite{Shi2000} as well as Quantum Monte Carlo (QMC) simulations  \cite{Ma2006, Sakhel2008}. Separation of species 1 and 2  is usually characterized in terms of a parameter $\Delta = U_{11} U_{22} - U_{12}^2$, defined in terms of the relative intraspecies ($U_{11}, U_{22}$) and interspecies  ($U_{12}$) interaction strengths. When $\Delta < 0$, so that particles of species 1 and 2 have a relatively strong repulsion, the system \changed{is predicted to phase separate}, \changed{whereas for $\Delta \geq 0$ the system should remain mixed} \cite{Ho1996, Ao1998, Timmermans1998,Pu1998, Esry1999, Trippenbach2000,Shi2000, Ma2006, Sakhel2008}. Recently, this \changed{criterion} has been verified in experiments with binary BECs \cite{Papp2008}. \changed{Earlier work on bosonic mixtures in the context of superfluid Helium have also predicted phase separation in the zero temperature limit for isotopes of different masses or concentrations  \cite{Chester1955, Miller1978}.} Indeed, all predictions of phase separation in binary \changed{mixtures of bosons} to date rely on a mismatch of interaction strengths or  of some other physical parameters on which the Hamiltonian depends (such as different particle masses, \changed{concentrations,} or external trapping potentials for each species). We refer to this scenario as \emph{de-mixing through interactions}.

In this article, we report the prediction of de-mixing in a binary mixture of bosons with identical masses and interactions, due a very different mechanism --- namely the effective attraction between indistinguishable bosons originating purely from quantum statistics.  \changed{This prediction is made for $\Delta = 0$ in contrast to earlier work for which this value would lead to a miscible system.} We refer to this \changed{scenario} as \emph{de-mixing through exchanges} or \emph{quantum de-mixing}, which occurs when the system kinetic energy is reduced by the formation of long exchanges of identical particles, leading to the spatial separation of the two components. 
This is, of course, the same mechanism underlying Bose-Einstein condensation, as first established by Feynman \cite{Feynman1953} and subsequently elaborated on \changed{\cite{Suto1993, Ueltschi2006}}. At low temperature, as the thermal wavelength becomes comparable to the interparticle distance,  quantum exchanges involving two or more {\it indistinguishable} particles become frequent, and \changed{condensation} sets in -- this effect is also responsible for phase separation. \\ % In particular, mean-field treatments \cite{Ho1996, Ao1998, Timmermans1998, Pu1998, Esry1999, Trippenbach2000} necessarily neglect correlations between particles --- and therefore also exchanges between indistinguishable particles --- and are therefore unable to describe this phenomenon. 
\\ \indent
For this study, we have elected to use the dipole-dipole interaction potential to describe the inter-- and intra--species interactions, for which the condition $\Delta = 0$ is always satisfied. The physics of ultracold dipolar bosons has fast become the subject of intense research activity. Bose-Einstein condensation of dipolar Chromium atoms has already been achieved \cite{Griesmaier2005}. The long-range and anisotropic nature of the interaction leads to many fascinating and novel phenomena \commentpj{(see the review \cite{Lahaye2009} and references contained therein)}.
\\ \indent
So far, there have been few calculations explicitly dealing with binary mixtures of dipolar bosons \cite{Goral2002,Saito2009}. In Ref. \cite{Goral2002}, the stability of a binary mixture with the components having oppositely oriented dipoles was investigated. In Ref. \cite{Saito2009}, spontaneous pattern formation associated with ferrofluidity was predicted and attributed to the anisotropic nature of the interaction. The role of finite range interactions in binary mixtures has also been addressed in previous studies \cite{Trippenbach2000, Alexandrov2002}. In particular, it was found that increasing the range of the interactions leads to increased mixing \cite{Trippenbach2000}.
\\ \indent
\section{Formulation}
We consider here a system comprising $N_a$ atoms of species $a$ and $N_b$ atoms of species $b$ confined in a harmonic trapping potential.  Let $M_a$, $M_b$ be the masses of each species, and $V_{aa}$, $V_{bb}$ and $V_{ab}$ be the inter- and intra-species interaction potentials. We consider here a two-dimensional confined geometry, with a transverse polarization field in the $z$ direction, for which the dipole-dipole interaction potential becomes isotropic and purely repulsive \footnote{This is \changed{in} fact a valid approximation for so-called \emph{pancake} traps --- highly anisotropic harmonic traps with transverse and planar trapping frequencies satisfying $\omega_z \gg \omega_{\rho}$. }, i.e.,
%\begin{eqnarray}
%V_{m m^{\prime}}({r}) = \frac{d_m d_m^{\prime}}{r^3} 
%\end{eqnarray}
\changed{$V_{m m^{\prime}}({r}) = d_m d_m^{\prime}/r^3$,}
between particles of species $m$ and $m^{\prime}$, at a distance $r$ from each other and with respective (electric or magnetic) dipole moments $d_m$ and $d_m^{\prime}$. Introducing a reference dipole moment $d_\circ$, we choose characteristic length and energy scales as $r_\circ = d_\circ^2 M_a/\hbar^2$ and $\epsilon_\circ = \hbar^2/(M_a r_\circ^2)$ respectively. 
The Hamiltonian for the two-component system is given, in dimensionless form, by:
\begin{eqnarray}\label{eq:hamcompute}
\hat{H} = &&  \sum_{i=1}^{N_a} \left( -\frac{1}{2} {\nabla}^2_{ai} + \Gamma {\bf r}^2_{ai} \right) + \sum_{i < j} \frac{\alpha^2}{|{\bf r}_{ai} -{\bf r}_{aj} |^3} \nonumber \\
&& +  \sum_{i=1}^{N_2} \left( - \frac{1}{2} \left(\frac{M_a}{M_b}\right) \tilde{\nabla}^2_{bi} + \Gamma \left(\frac{M_b}{M_a}\right){\bf r}^2_{bi} \right) \nonumber \\
&& + \sum_{i < j} \frac{\beta^2}{|  {\bf r}_{bi} -{\bf r}_{bj} |^3}
 + \sum_{i=1}^{N_a}  \sum_{j=1}^{N_b} \frac{\alpha \beta}{|{\bf r}_{ai} -{\bf r}_{bj} |^3}
\end{eqnarray}
where $\mathbf{r}_{m k}$ is the position of the $k$th particle of species $m$ and $\Gamma = 1/2 (L_{\rho} / r_\circ)^{-4}$ gives the trap strength,  $L_{\rho} = \sqrt{\hbar / M_a \omega}$ being  the harmonic oscillator length.
For brevity, we write the relative dipole amplitudes as $\alpha = d_a / d_0$ and  $\beta = d_b / d_0$. %so that our interaction potentials become $V_{aa} = \alpha^2 /r^3$, $V_{bb} = \beta^2 / r^3$ and $V_{ab} = \alpha \beta / r^3$. 
%--------------------  PLOTS ---------------------
\begin{figure}[t]
%\centering
\begin{minipage}[t]{\columnwidth}
\centering
\includegraphics[scale=1]{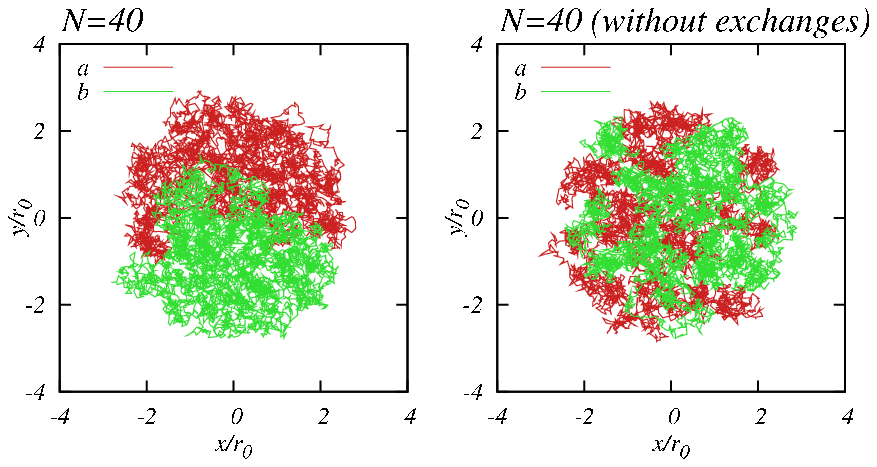}\vspace{2mm}
\end{minipage}
\hfill
\begin{minipage}[t]{\columnwidth}
\centering
\includegraphics[scale=1]{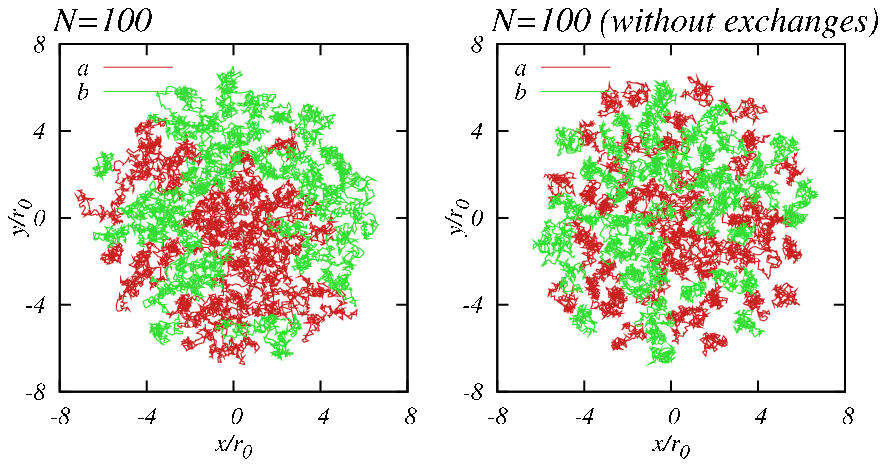}\vspace{2mm}
\end{minipage}
\vspace{-5mm}
\caption{\changed{(Color online) QMC configuration snapshots at temperature $T=0.5$ for species $a$ and $b$ in harmonic trap with $\alpha = \beta  = 1$. Snapshots show particle world lines.  In the upper plots $N=40$ and $\Gamma = 8$ whereas in the lower plots $N=100$ and $\Gamma = 0.5$.}}
\label{fig:snapshots}
\end{figure}

If by analogy with the case of hardcore boson mixtures, we define a mixing parameter $\Delta \sim V_{aa} V_{aa} - V_{ab}^2$, then  $\Delta = 0$ always. 

\section{Numerical results}

\changed{Henceforth, we choose $M_a = M_b$ and $N_a = N_b = N/2$.} We have investigated the finite temperature equilibrium properties of the system by  QMC simulations based on the Continuous-space Worm Algorithm \cite{Boninsegni2006,Boninsegni2006b}. This technique is numerically exact, to within a controllable statistical error.  We have carried out simulations with two values of $N$, namely 40 and 100, and with different values of the harmonic trap strength $\Gamma$, always chosen sufficiently small to keep the density in the middle of the trap below the crystallization threshold \cite{Pupillo06}. Our results are qualitatively the same for all cases considered.

\subsection{Identical and distinguishable species}

\begin{figure}[!t]
%\centering
\begin{minipage}[t]{\columnwidth}
\centering
\includegraphics[scale=1]{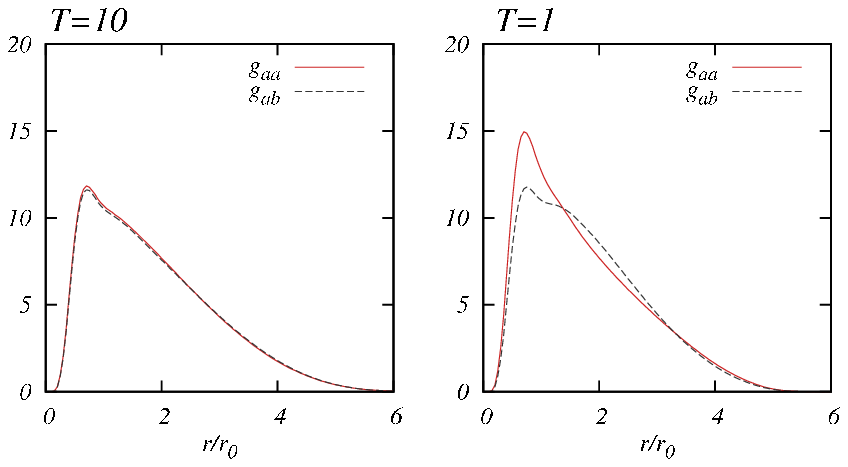}\vspace{2mm}
\end{minipage}
\hfill
\begin{minipage}[t]{\columnwidth}
\centering
\includegraphics[scale=1]{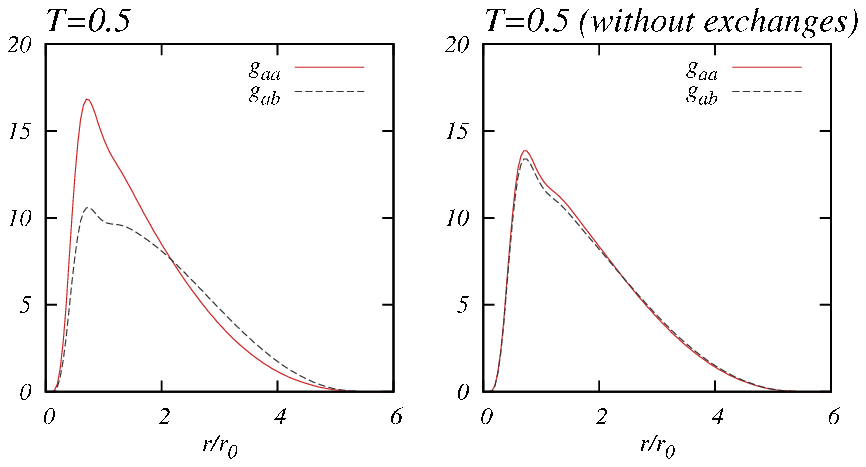}\vspace{2mm}
\end{minipage}
\vspace{-7mm}
\caption{\changed{(Color online) Integrated pair correlation function at temperatures $T=10, 1, 0.5$ between like ($g_{aa}$) and unlike ($g_{ab}$) particles for system in harmonic trap with $\Gamma = 8$, $\alpha^2 = \beta^2  = 1$ and $N = 40$.  Errors are $\lesssim 7 \times 10^{-2}$ in all cases. }}
\label{fig:grN40}
%\end{figure}
%\begin{figure}[!h]
%\centering
\vspace{4mm}
\begin{minipage}[t]{\columnwidth}
\centering
\includegraphics[scale=1]{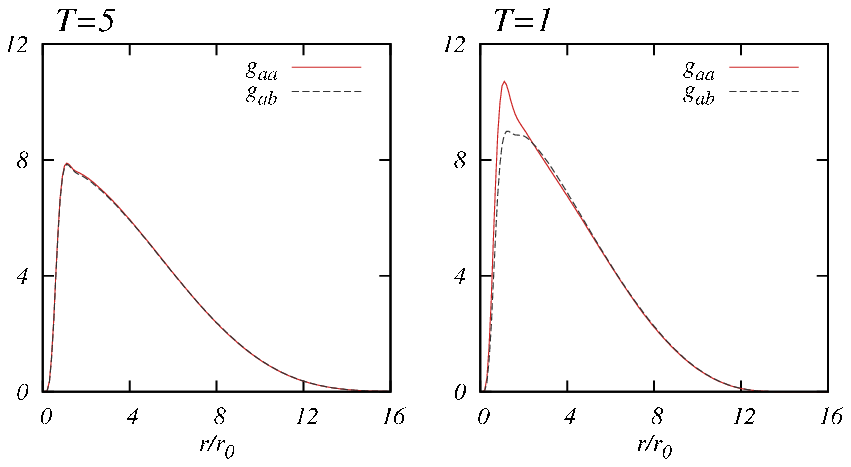}\vspace{2mm}
\end{minipage}
\hfill
\begin{minipage}[t]{\columnwidth}
\centering
\includegraphics[scale=1]{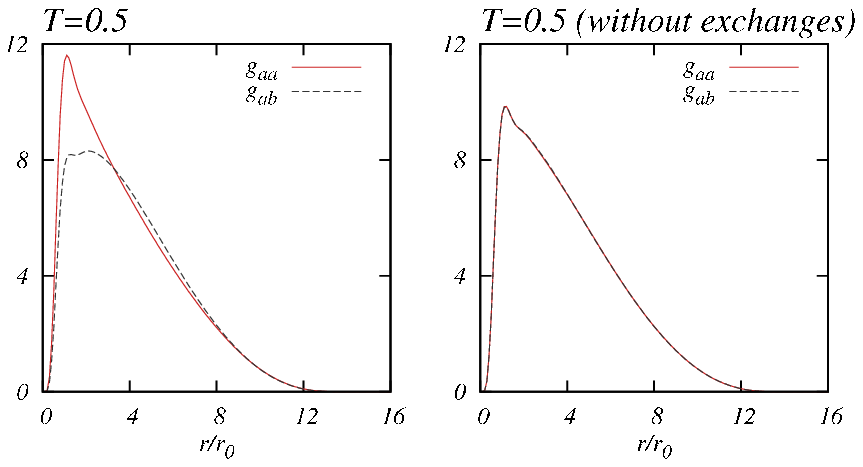}\vspace{2mm}
\end{minipage}
\vspace{-7mm}
\caption{\changed{(Color online) Integrated pair correlation function at temperatures $T=5, 1, 0.5$ between like ($g_{aa}$) and unlike ($g_{ab}$) particles for system in harmonic trap with $\Gamma = 0.5$, $\alpha^2 = \beta^2  = 1$ and $N = 100$.  Errors are $\lesssim 2 \times 10^{-2}$ in all cases. }}
\label{fig:grN100}
\end{figure}

We initially 
%investigate a small system size ($N = 40, \Gamma=8$) and 
set the dipole moments of each species equal (\changed{$\alpha = \beta = 1$}) so that the Hamiltonian (\ref{eq:hamcompute}) is symmetric to an interchange of labels $a$ and $b$. 
\changed{Typical QMC configuration snapshots are shown in Fig.~\ref{fig:snapshots} for $N=40$ with $\Gamma = 8$ (top) and $N=100$ with $\Gamma = 0.5$ (bottom). In each case, the left plot shows the case where exchanges between indistinguishable (like) particles are included, whereas the right plot shows the same system but without exchanges (ie. distinguishable particle statistics). For indistinguishable particle statistics like particles tend to aggregate, which is suggestive of de-mixing, whereas for distinguishable particle statistics the system remains mixed. Individual snapshots are not conclusive however, so this prediction is verified \commentpj{first in Fig.~\ref{fig:grN40} for the case $N=40$ and $\Gamma = 8$ and then} in Fig.~\ref{fig:grN100}  for \commentpj{a larger system size with} $N=100$ and $\Gamma = 0.5$. \commentpj{Both figures} show the {\em integrated} \footnote{Clearly, since the system is not translationally invariant one ought more properly look at the two-point correlation function $g_{m m^{\prime}}({\bf r},{\bf r}^\prime)$, with ${\bf r},{\bf r}^\prime$ measured with respect to the center of the trap. The function considered here is averaged over the whole trap; however, it still provides the quantitative information sought here.}  pair correlation function $g_{ab}(r)$,  which \commentpj{for the lowest temperature shown, $T=0.5$,} becomes suppressed at short distances with respect to $g_{aa}(r)$. That is,  the probability of finding \emph{unlike} particles separated by $r$ is less than that of finding \emph{like} particles at the same distance. Differently phrased, a particle of a given species is preferentially surrounded by like particles. }
%and represent what may be observable in single shot \emph{in situ} measurements with trapped ultracold atoms. density averages out, need to consider pair correlation function to quantify effect. \#\#\# }
\commentpj{For both system sizes, at} the highest temperature shown, there is no evidence of de-mixing, and the probability density of position for particles of either species only depends on the distance from the center of the trap. In this regime, entropy dominates.

As the temperature is lowered,  the two components separate. 
\changed{To verify that exchanges are responsible for the observed de-mixing \commentpj{Figs.~\ref{fig:grN40} (bottom) and ~\ref{fig:grN100} (bottom)  also show} the result} for $T$=0.5, but with exchanges turned off in the simulation (i.e., we regard particles as distinguishable). The resulting \commentpj{plots} shows the system is fully mixed. 
The energetic mechanism leading to phase segregation, is that particles can lower their kinetic energy by exchanging with like particles, thereby enhancing their spatial delocalization. \changed{We also expect de-mixing to occur for larger $N$ (relevant to experiments) as the relative interfacial energy decreases.}

%MB On second thoghts, maybe this can be stated here not as a footnote but part of regular text
\commentpj{It should be emphasized that the de-mixing predicted here is a finite temperature effect.}
Finally, on lowering the temperature even further, the system remixes. This effect, also observed in hardcore boson mixtures \cite{Ma2004}, 
%MBis due to the finite size of the trap. It 
occurs when the temperature $T$ is less than the level spacing $\hbar\omega = \sqrt {2\Gamma}$ of the confining harmonic potential. We have verified in our simulations that the remixing temperature decreases for smaller $\Gamma$.  
\\
\begin{figure}[tp]
\centering
\includegraphics[scale=1.]{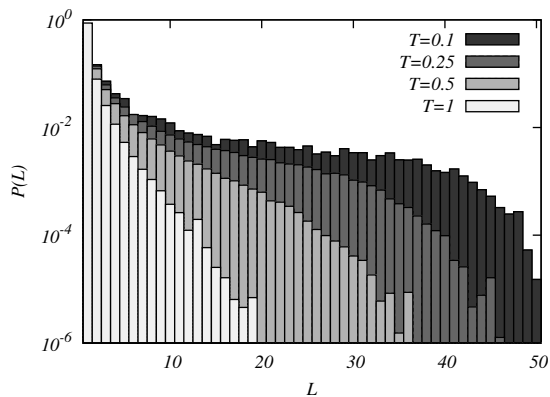}\vspace{-2mm}
\caption{Relative frequency of  permutation cycles of length $L$ for species $a$ with $N = 100$, at four different temperatures for the case where $\alpha^2 = \beta^2 = 1$. }
\label{fig:pcycles}
\end{figure}
\noindent
We further elaborate on the connection between the appearance of long permutation cycles and de-mixing referring to Fig.~\ref{fig:pcycles}, which shows the relative frequency of permutation cycles including different numbers $L$ of particles, for a single species. Note that, as $T \to 0$, the frequency of occurrence of cycles of permutation (ie. exchanges) involving almost all particles of each species (i.e., \changed{$L=50$}, in this case) increases dramatically. \changed{These exchanges are central to Bose-Einstein condensation, which in turn is responsible for the effective attraction between like atoms and therefore the observed phase separation.} %The observed phase separation due to long exchanges, is the main result of this Letter. 
Such quantum de-mixing is not particular to our choice of interaction. Indeed, we have verified that it also occurs for potentials of the form $V(r) \sim 1/r^{12}$ (and with $\Delta = 0$), which emulate the usual hard-core repulsion appropriate for ultracold alkali atoms. \changed{In the more general case where $\Delta > 0$ the repulsive interaction energy would be offset by the decrease of kinetic energy due to exchanges, although this effect may not be large enough to lead to phase demixing.}
\\ \indent
%Initially we will focus on two observables: first, the radial density profile, $\rho_m(r)$ (for species $m$), and the 
%(unnormalized) 
%pair correlation functions $g_{m m^{\prime}}(r)$, yielding the probability of finding two particles (of species $m$ and $m^{\prime}$) at a relative distance $r$.
%The spatial extent for which de-mixing occurs is given by the region in which $g_{ab} < g_{aa}, g_{bb}$. To determine the degree of de-mixing we introduce the parameter
%\begin{eqnarray} 
%\Omega = \frac{2 \max(g_{ab})}{\max(g_{aa})+\max(g_{bb})}.
%\end{eqnarray}
%Clearly when the system is fully mixed $g_{aa} = g_{bb} = g_{ab}$ so that $\Omega = 1$. Any deviation from from this indicates some degree of de-mixing has occurred. We plot $\Omega$ for $N=40, \Gamma=8$ and $N=100, \Gamma=0.5$ below. At high temperatures $\Omega \rightarrow 1$ indicating entropic mixing.  Conversely for both system sizes the de-mixing is greatest at $T \approx 0.5$. 
%\begin{figure}[tp]
%\centering
%\includegraphics[scale=1.2]{Figures_PDF/f3.pdf}
%\label{fig:subfig1}
%\caption{(Color online) The mixing parameter $\Omega$ as a function of temperature $T$ for two different systems: $N = 40$ with $\Gamma = 8$ and $N=100$ with $\Gamma = 0.5$. In both cases we have set $\alpha = \beta = 1$.}
%\end{figure}
%--------------------  PLOTS ---------------------
\begin{figure}[t]
\centering
\includegraphics[scale=1]{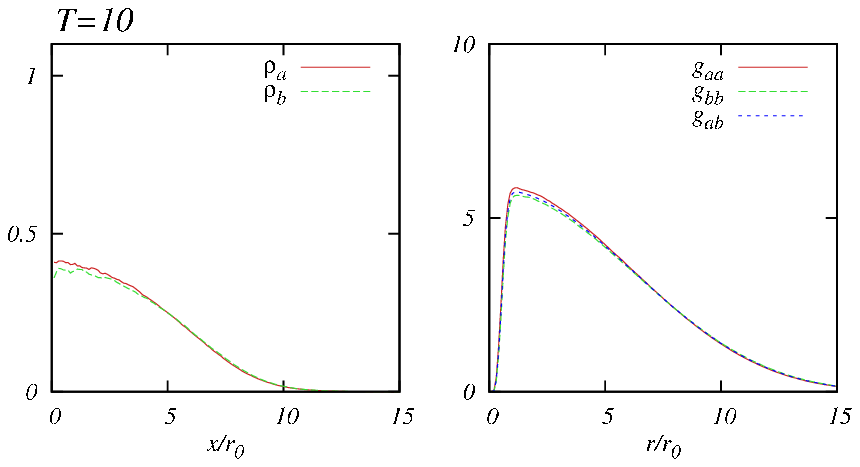}
\includegraphics[scale=1]{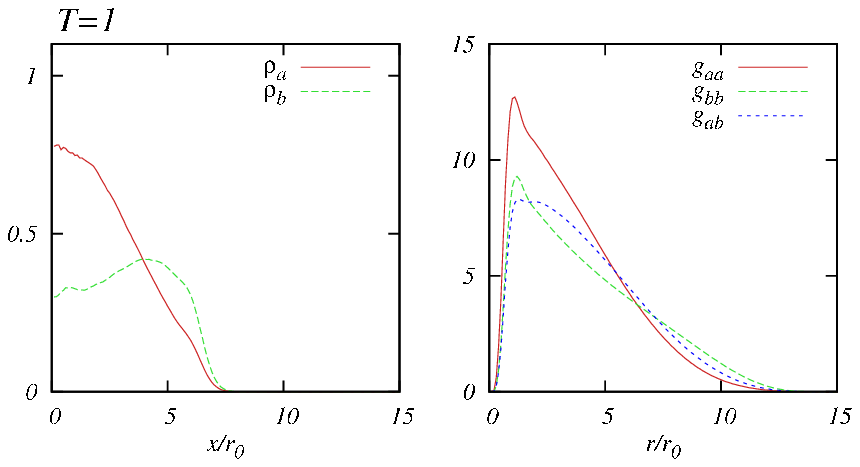}
\includegraphics[scale=1]{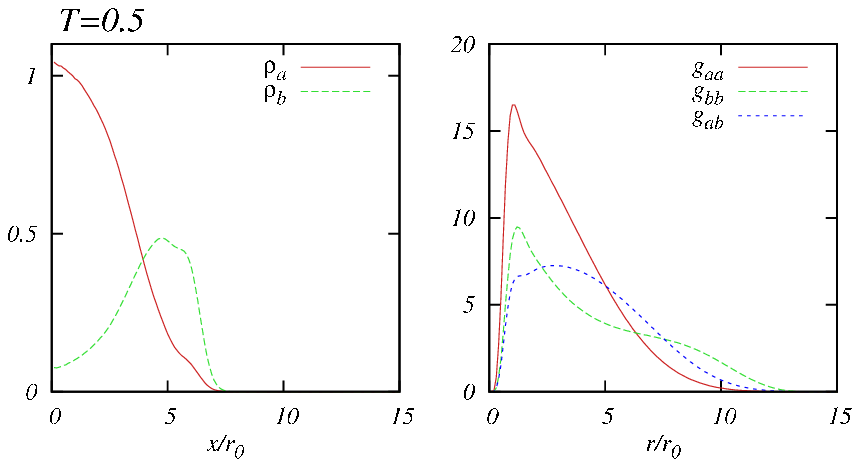}
\includegraphics[scale=1]{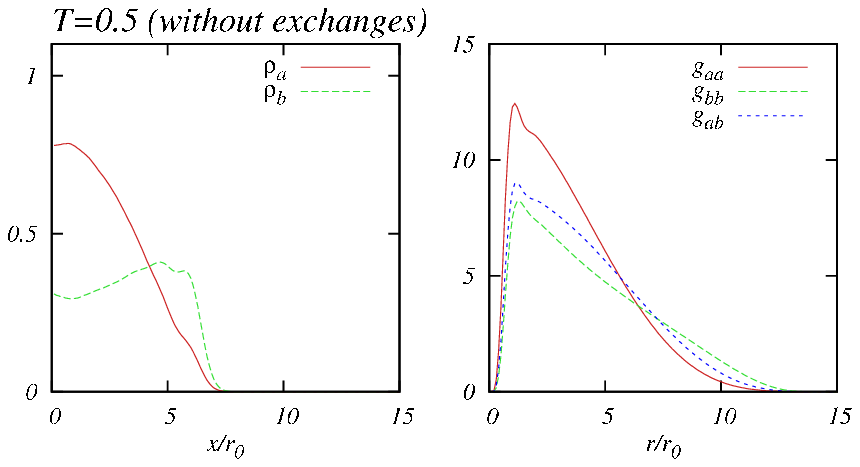}
\vspace{-2mm}
\caption{(Color online) Density profiles (left) and pair correlation functions (right) at temperatures $T=10, 1, 0.5$, for species $a$ and $b$ in harmonic trap with $\Gamma = 0.5$, $\alpha^2 = 1, \beta^2 = 1.21$ and $N = 100$. Also repeated for comparison is the case $T=0.5$ without exchanges (i.e., distinguishable quantum particles). \changed{Statistical errors are $\lesssim 2 \times 10^{-2}$ for all densities and $\lesssim 5 \times 10^{-2}$ for all pair correlation functions shown.}}
\label{fig:N100}
\end{figure}
%MB Hi Piyush, I have thought about it and because we cannot really ay anything certain as to what happens
%MB at T=0, I think we should leave it at that.
It is instructive to note that only a technique which {\it explicitly} treats exchanges of indistinguishable 
particles at finite temperature, can yield predictions of de-mixing such as those shown here, for components of identical masses and interactions.

\subsection{Non-identical and distinguishable species}

Next, we consider the situation where the dipole moments of each species are not equal (ie. $\alpha \neq \beta$). In this case, the symmetry of the Hamiltonian with respect of exchange of species $a$ and $b$ is explicitly broken, and we can expect phase separation to be evident in the radial density profiles $\rho_m(r)$, $m=a,b$, computed with respect to the center of the trap. 
\\ \indent
We consider for definiteness $\alpha < \beta$. In this case, species $b$ forms a shell around species $a$ due to the higher interaction energy of species $b$ in the presence of the harmonic trap. That is, the larger interparticle repulsion of species $b$ particles pushes that component to the outside of the trap. This can be seen in Fig.~\ref{fig:N100} for the case where $\alpha^2 = 1, \beta^2 = 1.21$. Specifically, as the temperature is lowered the partial overlap in density profiles between species $a$ and $b$ decreases, indicating the onset of de-mixing.  It is worth noting that for binary mixtures of non-identical species the presence of de-mixing is due to the combined effects of interactions and quantum exchanges. To verify that exchanges do indeed enhance de-mixing, we show in  Fig.~\ref{fig:N100} density profiles and pair correlation functions at $T=0.5$, for the simulations both with and without exchanges. 
\\ \indent
\commentpj{
The functions $\rho_m(r)$ can be used to quantify the degree of phase separation through the normalized overlap integral:
\begin{eqnarray}
\Lambda = \frac{\left[ \int \rho_a(r) \rho_b(r) \textrm{d} r \right]^2}{\left[ \int \rho_a(r)^2 \textrm{d} r \right] \left[ \int \rho_b(r)^2 \textrm{d} r \right] }.
\end{eqnarray}
When there is complete overlap ($\rho_a \propto \rho_b$) then $\Lambda = 1$ indicating total mixing, whereas for complete phase separation we have $\Lambda = 0$.   
\changed{Note that $\Lambda$ is not a good indicator of de-mixing when $\alpha = \beta$ since in this case the symmetry of the Hamiltonian means that the density profiles are identical when sufficient statistics are accumulated in the simulations. }
Fig.~\ref{fig:radialmixing} shows the radial mixing $\Lambda$ as a function of temperature for \changed{three} different combinations of dipole strengths ($\alpha$ and $\beta$). As the temperature decreases and/or the ratio between $\alpha$ and $\beta$ decreases phase separation becomes more pronounced. 
\begin{figure}[tp]
\centering
\includegraphics[scale=1.1]{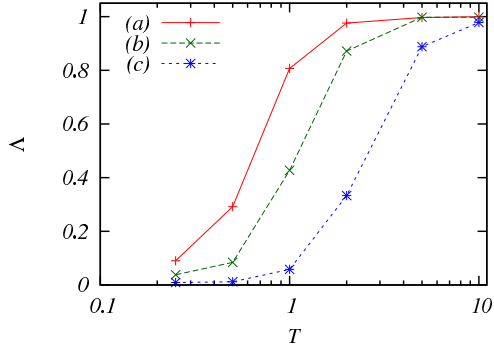}\vspace{-2mm}
\caption{(Color online) Radial mixing $\Lambda$ as a function of temperature $T$ for: (a) $\alpha^2 = 1, \beta^2 = 1.21$; (b) $\alpha^2 = 0.8, \beta^2 = 1.25$; and (c) $\alpha^2 = 0.5, \beta^2 = 2$.}
\label{fig:radialmixing}
\end{figure}
\\ \indent
}

\section{Conclusions}

In conclusion, low-temperature de-mixing purely due to Bose statistics is predicted to occur in a binary mixture of dipolar atoms, even when masses  and the dipole moments of each species are equal, i.e., inter- and intra-species interactions are identical. In the case where the dipole moments are unequal, interactions also contribute to de-mixing. As the temperature is raised, on the other hand, the system becomes miscible due to the entropy of mixing. The observation of the effect predicted here appears well within reach of current experimental efforts with cold dipolar systems. \changed{For our simulations the total peak density was $\rho \, r_o^2 \sim 1$, which for $^{52}\textrm{Cr}$ atoms gives $\rho \sim 10^{17}$ $\textrm{m}^{-2}$. This, while higher than those typical of magnetic traps, should be approachable using optical traps \cite{Pupillo06}. Moreover, optical traps allow for the simultaneous trapping of different hyperfine states \cite{StampurKurn1998}, which in principle allows for two distinct but equal mass species. The integrated pair correlation function should be measurable by averaging over several ``single shot'' absorption images \cite{Folling2005}.}\\

\section*{Acknowledgments}

This work was supported in part by the Natural Science and Engineering Research Council of Canada under
research grant G121210893, and by the Alberta Informatics Circle of Research Excellence. We are also grateful to F. Cinti for useful discussions. \\

\section*{References}

\bibliography{BinaryMixtures_long}

%\section*{References}
%\begin{thebibliography}{10}
%\bibitem{book1} Goosens M, Rahtz S and Mittelbach F 1997 {\it The \LaTeX\ Graphics Companion\/} 
%(Reading, MA: Addison-Wesley)
%\bibitem{eps} Reckdahl K 1997 {\it Using Imported Graphics in \LaTeX\ } (search CTAN for the file `epslatex.pdf')
%\end{thebibliography}

\end{document}